# Enhancement of Betatron radiation from laser-driven Ar clustering gas

L. M. Chen[1*], W. C. Yan[1], D. Z. Li[2], Z. D. Hu[1], L. Zhang[1], W. M. Wang[1], N. Hafz[3], J. Y. Mao[1], K. Huang[1], Y. Ma[1], J. R. Zhao[1,4], J. L. Ma[1], Y. T. Li[1], X. Lu[1], Z. M. Sheng[3], Z. Y. Wei[1], J. Zhang[1,3]

[1]*Beijing National Laboratory of Condensed Matter Physics, Institute of Physics, CAS, Beijing 100080, China*
[2]*Institute of High Energy Physics, CAS, Beijing 100049, China*
[3]*Department of Physics, Shanghai Jiao Tong University, Shanghai 200240, China*
[4]*Physics Department, Shandong Normal University, Jinan 250014, China*

Bright betatron x-ray has been generated using an Ar clustering gas jet target irradiated with a 3 TW ultra-high contrast laser. The measured emission flux with photon energy > 2.4 keV reaches $2 \times 10^8$ photons/shot. It is ten-fold enhancement comparing to the emission flux produced by using gas target in the same laser parameters. Observation shows that much larger electron beam charge and divergence angle lead to this improvement. Simulations point to the existence of cluster in gas results in the increasing of electron injection and much larger wiggling amplitude in wake-field, enriching the betatron x-ray photons.

Synchrotron light sources have proven very useful for a wide range of investigations. However, it is inappropriate for some applications because of its large cost and experimental footprint, relatively long pulse structure and synchronization difficulty in sub-picosecond pump-probe diagnostics. Hard x-ray emission from femtosecond (fs) laser-produced plasmas has been extensively studied in the past years, including for example refs. [1-3]. Such hard x-ray sources are of interest for a number of imaging applications [4, 5]. This kind of intense and ultrafast hard x-ray source can provide an alternative to synchrotron radiation sources due to its compactness, fs pulse duration and fs pump-probe ability, making it practical for widespread use in material and biological sciences.

At present, although greatly improved in source energy conversion efficiency and temporal duration [6, 7], most of laser-driven hard x-ray sources are spatially symmetric, except for high harmonic generation [8], Thomson scattering [9] and betatron radiation [10]. However, high harmonics and Thomson scattering are suffering from their low emission flux. A well collimated x-ray beam with fs duration was produced when a relativistic laser pulse propagates in an underdense plasmas: the electron injecting into plasma wakefield and experiencing periodically betatron transverse oscillation in the procedure of longitudinal acceleration [10]. Intense betatron radiation, with photon energy from keV x-rays [11] to gamma-rays [12], will be generated via the accelerated electrons undulating in the plasma channel. These oscillations occur at the betatron frequency $\omega_\beta=\omega_p/(2\gamma)^{1/2}$, where $\omega_p$ is the plasma frequency and $\gamma$ is the Lorentz factor of the electron beam [11]. It shows that characteristics of x-ray are mainly decided by parameters of electron during its acceleration, including the electron energy, beam charge and electron wiggling amplitude in wakefield. However, when using a gas target, improving accelerated electron energy and charge simultaneously is contradictory in some case, for example: low plasma density is necessary for longer pump-depletion length in gas target which leads to higher electron energy, but on the other hand, it also results in lower electron charge in self-injection. This is the reason why most of published electron charge using medium size laser facility is limited in the level lower than 100 pC. Some investigations tried to increase the injection wiggling amplitude using an asymmetric tilted pulse front [13] to enhance the betatron x-rays, but it usually effect the formation of wakefield in some case. Therefore, in order to enhance the betatron radiation, try to find an effective way to increase electron beam charge and electron wiggling amplitude in the wakefield are the subjects of many current investigations.

---

* Electronic address: lmchen@aphy.iphy.ac.cn

In this Letter, we present generation of large charge electron beam as well as strong betatron x-ray using an Ar clustering gas jet target irradiated with a 3 TW femtosecond laser. The measured emission flux with photon energy > 2.4 keV reaches $2\times10^8$ photons/shot, which is ten-fold enhancement comparing to the emission flux produced by using gas target in the same laser parameters. Experiment and simulations point to the existence of cluster in gas background results in the increasing of electron injection and much larger wiggling amplitude in the acceleration wake-field, enriching the betatron x-ray photons.

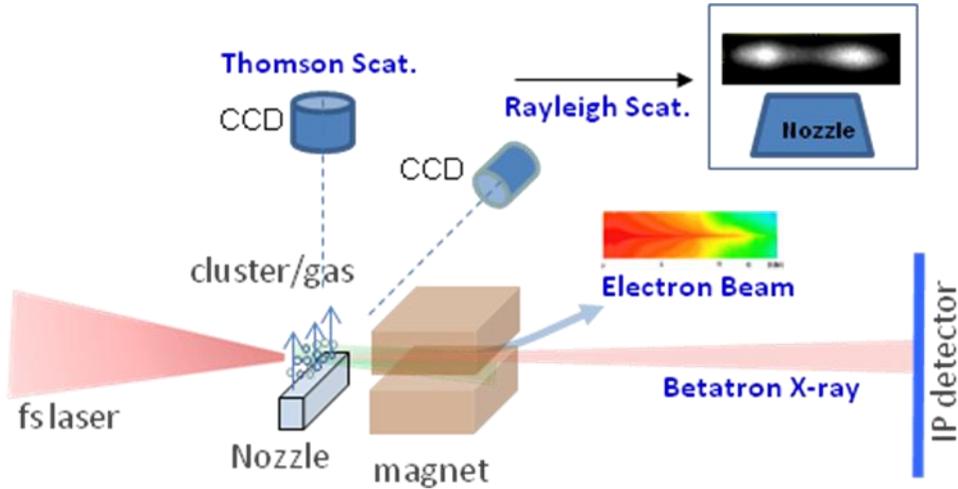

Figure 1: (color online) Schematic of experimental setup. The upper right inset shows the Rayleigh scattering measurement.

The experiments are carried out by using the XL-II laser facility in the Institute of Physics, which includes a 10 Hz, 300 mJ Ti:Sapphire laser working at the centre wavelength $\lambda$ = 800 nm. The pulse with duration $\tau_0$ = 80 fs is focused with an f /6 off-axis parabola (OAP) onto a focal spot of size $w_0$ = 4 μm. In the focal region the laser average intensity is I = $3.0\times10^{18}$ W/cm$^2$. With the help of optical parametric chirped pulse amplification, the laser pulse contrast, compared to its ns prepulse, has been improved to $10^9$. As shown in **Fig. 1**, a supersonic pulsed gas (Ar) jet is used, which is a 1.2 mm × 10 mm rectangular nozzle at the exit. The laser pulses incident in the center of 10 mm width and along the 1.2 mm length. A strong magnet is placed after the nozzle to disperse electrons, an imaging plate follows the magnet to record electron beam. A filtered big size imaging plate is located in the laser propagation direction to detect the x-ray flux and spatial profile with photon energy > 2.4 keV. Sometimes, several filters will be placed in front of IP to selectively attenuate the x-ray flux and estimate the spectral content as Ross filters, which is correlated with the electron spectrum. A knife edge is introduced to measure the x-ray source size. A probe beam is used for detecting the shadowgraph. The cluster size is estimated by employing the Hagena scaling law [14]. An average size of ~8 nm in diameter is assumed at a stagnation pressure of 4 Mpa, with a surrounding gas density $2\times10^{19}$ cm$^{-3}$. Cluster spatial distribution profile along laser propagation direction was measured by employing Rayleigh scattering diagnostics. As seen in the upper right inset of **Fig. 1**, along the laser propagation direction, there is a bright scattering region followed by a hollow. It means the cluster is mainly generated on the edge of nozzle where a gas shock is benefit for gathering gas atom to form clusters. The dark part in the nozzle middle shows no cluster but gas atoms exists.

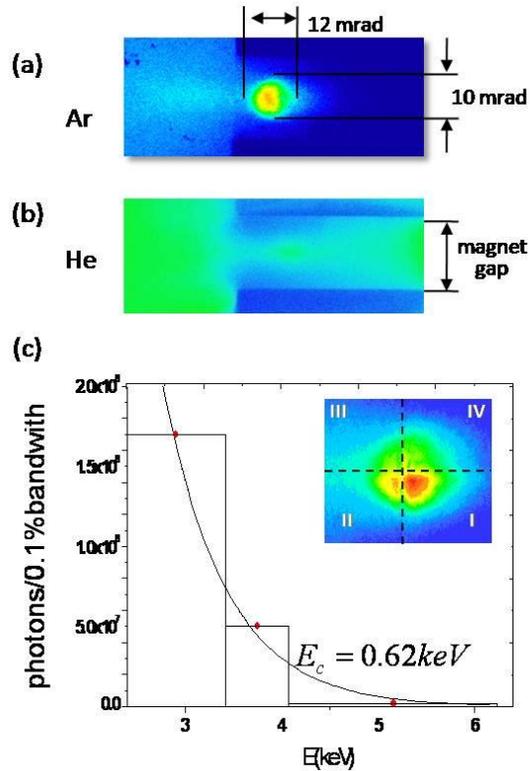

Figure 2: (color online) Betatron emission characteristics: The emission beam profile obtained using 36 μm Al foil wrapped IP for Ar target (a) and He target (b) respectively. Spectrum is shown in (c). The inset shows the beam profile recorded after Ross filters 18 μm Al (I), 18 μm Al + 10 μm Cu (II), 18 μm Al + 20 μm Cu (III) and 18 μm Al + 30 μm Cu (IV) respectively, forming energy region 2.4-3.4 keV, 3.4-4.1 keV and 4.1-6.2 keV.

By using Ar clustering gas as an acceleration media, the photon flux of fs laser driven betatron radiation is greatly enhanced in our experiment. **Fig. 2a** shows the Ar betatron x-ray profile obtained in a single laser shot with photon energy > 2.4 keV. Compared to the x-ray profile from a gas He target in same laser parameters, the Ar betatron emission shows much more intense. The emission flux reaches to $2\times10^8$ for a single shot with a divergence angle ~10 mrad, whereas it is only ~$1\times10^7$ in the case of a He gas target, as shown in **Fig. 2b**. Compared to previous observations when gas target was used, the betatron x-ray flux we obtained is over 1000 times higher than the case in the similar laser power [10], and it is also four-fold increasing to the case when 70 TW laser was used [11]. **Fig. 2c** shows the spectrum obtained by using Ross filter technique. Most of betatron x-ray photons is concentrated in the energy range of 2.4-3.4 keV and expands to over 6.2 keV. Fitting with a synchrotron distribution as defined in Ref. [15], the critical energy we obtained is $E_c=\hbar\omega_c=0.62$ keV. It should be emphasized that the betatron flux depends critically on the laser contrast. The flux is reduced by two orders of magnitude if the laser pulse contrast decreases from $10^9$ to $10^7$ with constant laser pulse energy. Pre-expansion of a solid-density cluster by the laser pre-pulse leads to ineffective electron heating and wakefield injecting, as described below, and results in the decrease of x-ray flux in this case.

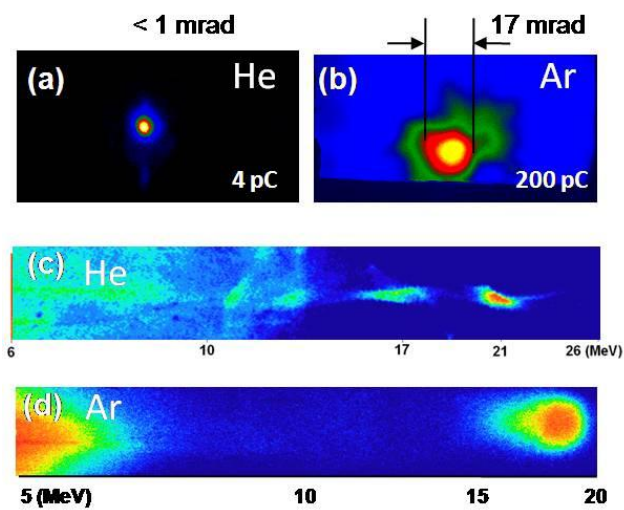

Accelerated electron spectrum was measured simultaneously because it is reflecting the behavior of electron in wiggling which leads to betatron x-ray generation. As shown in **Fig. 3a and 3b**, the electron beam profile measured using an Al foil wrapped imaging plate directly after the nozzle (without magnet) in the case of Ar cluster target and He gas target. The beam charge reaches to be 200 pC with electron energy higher than 1 MeV. It is over 50 times higher than the case of He gas in the same density was used. We noticed the divergence angle of electron beam using Ar cluster target, ~ 17 mrad, is much larger than the case of He gas target which is as small as 1 mrad. It reveals that electron experiences much larger oscillation during accelerating. **Fig. 3c and 3d** show the spectra obtained in the same interacting condition. Quasi-monoenergetic electron beam is obtained in both cases with almost similar electron energy. It is 18 MeV for Ar and 22 MeV for He. However, the beam charge and beam divergence angle in the case of Ar cluster are much larger than the case of He gas target. This measurement is consistence with beam profile results. It shows the cluster gas target is more suitable to get high charge electron beam acceleration, than the gas target, with larger divergence angle. We use these electron parameters to estimate the betatron x-ray radiation. The critical energy $E_c[\text{keV}]=5.3\times10^{-24}\gamma^2 n_e[\text{cm}^{-3}]r_0[\mu m]=0.53$ keV, which is similar to our experimental fitting above. The plasma wiggler strength $K=\gamma k_b r_0=14.1$ for photons with energy > 2.4 keV. For a betatron wavelength $\lambda_b=(2\gamma)^{1/2}\lambda_p=44$ μm, the number of betatron oscillation executed by a electron for laser-produced ion channel $N_0=cT_{int}/\lambda_b=22$. So, the total betatron flux with energy of 2.5 keV is: $N_x\approx 4.4\times10^{-4}N_e N_0 K=1.7\times10^8$. This value is in reasonable agreement with experimental detection.

Figure 3: (color online) Electron beam characteristics: The beam profile obtained using 18 μm Al foil wrapped IP for He target (a) and Ar target (b) respectively in the same laser parameters. The electron spectra are shown in (c) and (d) which obtained after the dispersive magnet.

Simulations using a 2D fully electromagnetic particle in cell (PIC) code have been performed, where a linearly polarized laser pulse with $\sin^2$ pulse envelope is launched along the +x direction onto clustering gas region with 10 nm diameter and $0.1n_{cr}$ (critical density) clusters and $2\times10^{18}$ cm$^{-3}$ for surrounding gas density. **Fig. 4a and 4c** show a snapshot of the electron distribution profile at t=1.08 ps in $I_{peak}=1\times10^{19}$ W/cm$^2$ in the case of pure uniform electron plasma and in the case of cluster dotted in gas background.

As shown in **Fig. 4a**, small amount of electrons injected into the second and following wakefields and it undergoes small oscillation amplitude in the acceleration procedure. Several electron beams with different energy will be obtained as seen in **Fig. 3c**. However, when a region of clustering gas target exist before the following gas target, which is same as our Rayleigh scattering measurement shown in **Fig. 1**, a large quantity of electrons inject into the first wakefield and the first bubble becomes much wider in temporal, see in **Fig. 4c**. During acceleration, two bunches of electron symmetrically experience much larger wiggling amplitude (~8λ) in the plasma wakefield for a long distance. The energy spectra ($E_x$, t) at the same time interval are shown in **Fig. 4b and 4d** for the case of pure gas and the case of clusters in front of gas target. In case of pure gas, there is no electrons in the first wakefield and electron charge injected into following wakefields as low as 36 pC in total. However, in case of Ar clustering gas, huge quantity of electrons injected into the first wakefield. The total electron charge is about 144 pC.

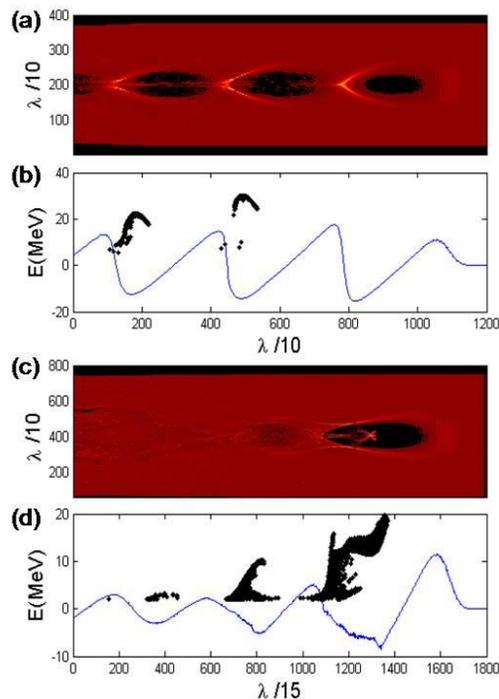

Figure 4: (color online) Simulation 2D snapshots of the electron density distribution of gas and (a) clusters (c) at time (with unit of laser optical cycle) of t=1.08 ps. The energy spectra are dependence of plasma location at the same time interval are shown of gas (b) and clusters (d).

We have to mention that effective electron acceleration as well as betatron production occurs only in the case of clustering gas mixture target. As described in ref. [16], if using only pure and large size (~μm) clusters (without the surrounding gas), no quasi-monoenergetic electron beam can be obtained and the plasma wake will be destroyed and broken into many filaments by those density un-uniform clusters. However, in our experimental condition, the cluster size is in nm scale and full of background gas atoms, which ensure the stimulation of wake field under intense laser irradiation.

In summary, we have presented a method for generation of intense betatron x-rays using a clustering gas target irradiated with a 3 TW ultra-high contrast laser. The intensity of the Ar x-ray emission has been measured to be $2 \times 10^8$ photons/pulse, which is ten-fold enhancement comparing to the emission flux produced by using gas target in the same laser parameters. Simulations point to the existence of cluster results in the increasing of electron injection and much larger wiggling amplitude in wake-field, enriching the betatron x-ray photons. Here in this regime, electrons are efficiently driven in a 10 fs time scale, producing enhanced x-ray emission with duration about three orders shorter than that of typical pulses produced by synchrotron sources. Together with the assuming source size of plasma wavelength ~10 μm, the peak brightness of the radiation in energy range about 3 keV is estimated to be ~$5 \times 10^{21}$ photons/s/mm$^2$/mrad$^2$/0.1%BW, which is comparable to the peak brightness of the third generation synchrotron radiation sources. This ultra-intense, table-top synchrotron with fs duration may make possible for "single-shot" laser-driven x-ray ultrafast pump-probe applications.

We thank L. T. Hudson and J. F. Seely for fruitful discussions. This work was supported by the NSFC (Grant No. 60878014, 10974249, 10735050, 10925421 and 10734130).